\documentstyle [12pt,aasms]{article}

\begin{document}

\title{The Effects of Substructure on Galaxy Cluster Mass Determinations}
\author{Christina M. Bird}
\affil{Department of Physics and Astronomy, University of Kansas, Lawrence,
KS 66045}
\affil{tbird@kula.phsx.ukans.edu}
\vskip 4cm
\noindent{Accepted for publication in the {\it Astrophysical Journal Letters}}

\begin{abstract}

Although numerous studies of individual galaxy clusters have demonstrated
the presence of significant substructure,
previous studies of the distribution of masses of galaxy clusters
determined from optical observations have failed to explicitly correct for
substructure in those systems.  In this {\it Letter} I present the
distributions of velocity dispersion, mean separation, and
dynamical masses of clusters when substructure is eliminated from the
cluster datasets.  I also discuss the changes in these distributions because
of the substructure correction.
Comparing the masses of
clusters with central galaxies
before and after correction for the presence of
substructure reveals a significant change.
This change is driven by reductions in the mean separation of galaxies,
not by a decrease in the velocity dispersions as has generally been
assumed.  Correction for substructure reduces most significantly the masses
of systems with cool X-ray temperatures, suggesting that
the use of a constant linear radius (1.5$h _{100} ^{-1}$ Mpc in this study)
to determine cluster membership is inappropriate for clusters spanning
a range of temperatures and/or morphologies.

\end{abstract}

\section{Introduction}

Studies of galaxy cluster dynamics, simulations of structure formation and
analytical arguments suggest that we are living in the era of cluster
formation.  For this reason, the assumption of dynamical equilibrium made
in early cluster studies no longer seems valid.  Over the last 15 years,
the study of substructure in clusters
has become vital to the study of clusters in general.  From the
cosmological point of view, clusters are useful for tracing the large-scale
distribution of matter in the Universe, for determining the value of the Hubble
constant through the Sun'yaev-Zel'dovich effect and other methods,
and for measuring the
value of the matter density of the Universe.
The presence of substructure implies
that non-equilibrium models need to be incorporated into these studies
to correctly replicate the behavior of the observed Universe.  Similarly,
the density of a galaxy's environment appears to have profound effects on
the structure and evolution of galaxies.  A high frequency of substructure
in clusters implies that the environment of any particular galaxy needs to
be determined locally, not globally (as first pointed out by Dressler, 1980).

A wide variety of observations in both the optical and X-ray regimes
provide convincing evidence that substructure in clusters is common,
if not ubiquitous.  X-ray imaging observations with the {\it Einstein}
satellite first revealed that complexity in the intracluster medium (ICM)
is frequent (Forman et al.\ 1981;  Henry et al.\ 1981), in contrast to
the smooth configurations assumed in early studies (cf.\ Kent \& Gunn 1982;
Kent \& Sargent 1983).  Since then, both
optical (Geller \& Beers 1982; Baier 1983; Beers et al.\ 1991; Bird 1993,
1994a,b) and X-ray studies (Jones \& Forman 1992; Davis \& Mushotzky 1993;
Mohr, Fabricant \& Geller 1993) have revealed the presence of significant
substructure in galaxy clusters.

The physical importance
of substructure is less clear.  Although most substructure diagnostics
are sensitive to deviations on small mass scales (Escalera et al.\
(1994) claimed that most of the substructure they detect is on the order
of 10\% of the total cluster mass), careful analysis reveals that
kinematical and dynamical estimates of cluster properties can be severely
affected by these apparently low levels of contamination (cf.\ Beers,
Flynn \& Gebhardt 1990; Bird 1994a).

In this paper I present an analysis of the dynamical masses of rich galaxy
clusters based on the velocities and positions of cluster galaxies.
The cluster dataset consists of clusters
with central dominant galaxies which have at least 50 measured redshifts
(the redshifts are taken from the literature, listed in
Table 1 of Bird 1994a).  These clusters are limited to morphological
types cD (optical) or XD (X-ray); the selection criteria are discussed in
more detail in Bird 1994a.

The large observational database permits use of an objective partitioning
algorithm called KMM (Ashman, Bird \& Zepf 1994) to identify and eliminate
substructure ``contamination'' from the primary subclusters, those that
contain the central galaxy.  The optical data is supplemented by X-ray
temperatures from the literature; these values are published in Bird,
Mushotzky \&
Metzler (1995) along with their sources.
The primary purpose of the current work is to show that even in these clusters,
long believed to be the most
dynamically-evolved systems, the distribution of observed masses is
dramatically changed if substructure is objectively eliminated from the
cluster datasets.  This result is in contrast to that of Biviano et al.\
(1993) and Escalera et al.\ (1994),
who claimed that the presence of substructure did not affect the
results of their optical mass determinations.

In Section 2, I review the dynamical mass estimators
and provide the distributions of velocity dispersion, mean galaxy
separation and dynamical mass for the cluster dataset, both before and
after substructure correction. In addition I quantify the effects of
substructure correction on these distributions.  I discuss various causes
of the change in the cluster mass distribution in Section 3.

\newpage

\section{Dynamical Mass Estimators}

For a system in dynamical equilibrium, the Jeans
equation relates the kinetic energy of the galaxies to the binding mass of
the cluster:
\begin{equation}
- {{G~n_{gal} M_{opt}(r)} \over r^2} = { d(n_{gal} \sigma ^2 _r) \over dr} +{2
n_{gal} \over { r \sigma _r ^2} } (1 - \sigma _r ^2 / \sigma ^2 _t)
\end{equation}
(Merritt 1987), where $M_{opt}$ is the optically-determined binding mass,
$n_{gal}$ is the galaxy number density, and $\sigma _r$ and $\sigma _t$ are the
radial and tangential velocity dispersions respectively.
By taking the fourth moment of the Jeans equation, Heisler, Tremaine \&
Bahcall (1985) derive orbital constants for the {\it projected mass
estimator}:
\begin{equation}
 M_p = {{\xi} \over {GN}} \sum _i v_i^2  R_i
\end{equation}
where $R_i$ is the projected distance of galaxy $i$ from the D/cD
galaxy, and $v_i$ is the velocity of galaxy $i$ with respect to the robust
estimator of the velocity
location of the cluster, $C_{BI}$ (see Beers, Flynn \& Gebhardt 1990 for a
complete discussion of the application of robust estimators).
The factor $\xi$ is equal to ${32}
\over \pi$, assuming a distribution of tracer particles with
isotropic orbits moving in a smoothly-distributed gravitational potential.

The assumption of isotropic orbits has not
been well-tested, although preliminary X-ray masses from {\it ROSAT} and
{\it ASCA} suggest that at least in a few clusters, the assumption of
isotropy in the galaxy orbits is consistent with the observed data
(Mushotzky, 1995).
The isotropic orbital constant has been used in most other studies
and more importantly, permits a
direct comparison with virial theorem mass estimates
(for which the assumption of an isotropic orbital distribution is necessary
but not explicit in the traditional formulation; see Heisler, Tremaine \&
Bahcall 1985).
The projected mass estimator is statistically preferable to the
virial mass estimator, especially for small datasets,
and in the majority of clusters yields masses which
are consistent with the virial estimator
within the errors (Postman, Geller \& Huchra 1988).

To correct for the presence of substructure in the cluster datasets, I have
used the KMM mixture-modelling algorithm (McLachlan \& Basford 1988).
KMM is an implementation of a maximum-likelihood technique which assigns
each galaxy into a prospective parent population, and evaluates the improvement
in fitting a multiple-group model over a single-group model.
KMM may be applied to data of any dimensionality;
for the cluster data I have simultaneously partitioned the velocity and
galaxy position data.
In addition to its use in the current study, it has been used as a hypothesis
test for the detection of bimodality (Ashman, Bird \& Zepf 1994, and
references therein).  See Bird (1994a) for details of the partitioning.

In any objective partitioning algorithm, verification of the partition
is the most difficult and subjective part.  For the cluster substructure
partitions, we have used several techniques to verify that the KMM algorithm
is behaving sensibly.  These include comparison with independent X-ray
images (Davis et al.\ 1995; Bird, Davis \& Beers 1995) and comparison with
the substructure allocation determined by other authors for their optical
data (especially Malumuth et al.\ 1992 and Pinkney et al.\ 1993).  In all
cases, the structures identified by independent methods verified the
objects identified by KMM.

In Table 1, I present $M_p$ for the clusters in the limited
sample defined above, as well as the robust estimator of the
velocity dispersion $S_{BI}$ (Beers, Flynn \& Gebhardt 1990) and
the mean distance of member galaxies from the cluster centroid, $<r _{\bot}>$.
These are determined within a radius of
1.5$h_{100} ^{-1}$ Mpc, one Abell radius (the effects of this ``fixed
aperture'' calculation are discussed below).  Use of the Abell radius
as a ``fixed aperture''
for mass determinations is common; see Biviano et al.\ (1993) and Beers
et al.\ 1995 for examples.  The position of the central
galaxy is used as the cluster centroid, following Beers \& Tonry (1986).
For the clusters in this sample which are included in the Beers \& Tonry
list, I have verified that the cD and X-ray centroids agree to within
1 arcminute, the pointing accuracy of the {\it Einstein} X-ray images.
Unprimed quantities are not corrected for
the presence of substructure; primed quantities include elimination of
galaxies identified by the KMM algorithm as contaminating.

It is important
to note that some of these ``contaminating'' structures are probably
themselves gravitationally bound, distinct subgroups in the potential of
the central galaxy host subcluster.  However,
it seems reasonable that low-mass systems undergoing
mergers with more massive clusters will be severely disrupted during the
interaction.  It is inappropriate to apply the dynamical mass
estimators, which are predicated on the systems under study
being steady state, to the ``contaminating'' structures, because we have no
objective way to distinguish which systems are disturbed and which are not.
The central galaxy
provides a useful tool by which we can objectively identify the
primary cluster.  This approach to mass correction differs from that of
Biviano et al.\ (1993), who claimed that substructure did not introduce
significant uncertainties into their work, and Escalera et al.\ (1994),
who assumed that all contaminating substructures were gravitationally bound
and undisrupted.

In Figure 1, I present the distributions of velocity dispersion, mean
distance of a cluster galaxy from the cluster position centroid, and $M_p$.
The visual impression provided by these histograms
suggests that the distribution of velocity dispersions has not changed
significantly, but that the other distributions have.  This subjective
impression can be quantified through use of a two-distribution
Kolmogorov-Smirnov test (Press et al.\ 1988).
The KS test is useful for distinguishing between
parent populations of observational distributions (although it cannot be
used to demonstrate that two distributions are the same).  The distributions
of velocity dispersion $S_{BI}$ are consistent with being drawn from the
same parent populations.
The before-and-after
distributions of $<r _{\bot}>$ and
$M_{p}$ are strongly inconsistent with each other (at a significance
level of 1\% in each case).

In Figure 2, I present the cumulative distributions of $M_p$.  The dotted
line represents the distribution without substructure correction; the solid
line represents the corrected distribution.  Note that although the shape
of the histograms in Figure 1 has changed substantially, the shape of the
cumulative distributions (their slope) is not severely affected by the
substructure correction.  The normalization, however, is reduced after the
inclusion of substructure in the analysis.

Note that contrary to previous impressions, overestimates of dynamical
masses do not appear to be caused by overestimates of the cluster velocity
dispersions, as is commonly assumed.  In any individual galaxy
cluster, the velocity dispersion may increase or decrease after KMM is
used to eliminate substructure.  On the other hand,
the mean distance parameter decreases
in almost every system in the limited cluster sample.  To some extent,
this may merely indicate that the use of the ``3-$\sigma$'' velocity filter
eliminates more line-of-sight structure than use of a fixed aperture radial
cut-off does projected structure.  However, it may also suggest that
overestimates in dynamical mass are due to the inclusion of
galaxies which are not within the virialized core of the cluster.
I test this possibility in the next section.

\section{What's Going On?}

Ashman (1992) argues that for a system to be virialized by the present epoch,
it must have a density of 5.8 $\times 10^{13}$ M$_{\odot}$
Mpc$^{-3} \equiv \rho _{vir}$ (see also Peebles 1993).
If we assume that all clusters have
similar formation epochs, and
that this value is a minimum, we can define a relationship
between density, radius and the depth of the gravitational potential well
of the cluster.  This depth may be estimated using either the velocity
dispersion of the cluster galaxies or the temperature of the X-ray
emitting gas (cf.\ Sarazin 1988); here we will use temperature, because it is
unaffected by the substructure corrections applied to the optical data.
The proportionality depends on the galaxy cluster being
in hydrostatic equilibrium within the core and gravity being the only
source of energy for either the galaxies or the gas, in which case:
\begin{equation}
T_X \propto {M \over r_{vir}} \propto {\rho _{vir} r_{vir}^3 \over
r_{vir}} \propto r_{vir}^2
\end{equation}
Therefore we can define the {\it radius of virialization} $r_{vir}$,
the distance to which the cluster is expected to have reached dynamical
equilibrium at the present epoch,
by scaling to X-ray temperature $T_X$.  It is convenient to use
the values for Coma, which has a gas temperature of 8.4$^{+1.1}_{-0.9}$ keV
(Watt et al.\ 1992) and $r_{vir} \approx r_A$, an Abell radius (The \& White
1986; Evrard, 1994, private communication).  While defining a physical
quantity on the basis of one cluster (which does not itself meet the
morphological criteria for membership in the current sample) is clearly
less than ideal, Coma's X-ray and optical properties are typical of rich
clusters and probably do not introduce a large uncertainty in this
argument.

Using the values of X-ray temperature for the 21 clusters in the limited
sample which have reliable X-ray observations (Bird, Mushotzky
\& Metzler 1995),
we find that the range of radius included by the 1.5 Mpc cutoff is
$1-2.5r_{vir}$.  For the coolest clusters
(A194, A1060, A2052, A2063, A2634, A2670, A3558 and DC1842-63, all with $T <
4$ keV),
using a 1.5 Mpc cutoff radius samples portions
of the cluster environment well outside the region expected to be virialized.
The remainder of the clusters in the limited sample have temperatures
between 6 and 9 keV, and values of $r_{vir}$ similar to 1.5 Mpc.
The last three columns of Table 1 provide $r_{vir}$, $M_p (< r_{vir})$
and $M_p ^{\prime}(< r_{vir})$ for the clusters with X-ray temperature
determinations.  $M_p (< r_{vir})$ and $M_p ^{\prime}(< r_{vir})$ are the
projected mass within $r _{vir}$ before and after substructure correction,
respectively.

The behaviour of the ``before-and-after'' distributions of velocity
dispersion, mean separation and dynamical mass for the 1.5$h^{-1}$ Mpc cutoff
helps to quantify this effect.
If the eight coolest clusters are removed,
a two-distribution KS fails to distinguish between {\it any} of the pairs of
distributions.  That is, the dramatic change in the distributions of mean
separation and dynamical mass quantified in Section 2 is due to changes in
only the clusters with the lowest X-ray temperatures (and presumably the
most shallow gravitational potentials).  If we consider the distributions of
dynamical mass calculated within $r_{vir}$ rather than within a fixed
aperture, the correction for substructure no longer significantly changes
the distribution of masses (the two-distribution KS test has a significance
level of 27.5\%).
Even if the assumptions made in the estimate
of $r_{vir}$ are incorrect -- if clusters don't all form at the same epoch
or if their densities are much different than $\rho _{vir}$ -- this
result suggests that rather than using a fixed aperture for mass determinations
(see, for instance, Biviano et al.\ 1993), use of a physically-motivated
radius for each cluster makes the most efficient use of the data and
reduces uncertainties due to the presence of substructure.
This result depends only weakly on the use of the Coma cluster to define
$r _{vir}$.  Note however
that even within $r_{vir}$, the substructure correction can significantly
affect the dynamical mass estimators for any particular cluster,
as is the case for A2634.

There are a couple of reasons why the present result differs from the
earlier work of Biviano et al.\ (1993) and Escalera et al.\ (1994).
Biviano et al.\ limit their discussion to masses determined within 0.75$h^{-1}$
Mpc of the cluster centroid, in order to reduce the effects of substructure
in their mass determinations.  This approach works but fails to make the
most efficient use of the data; as this paper shows, for the galaxy clusters
with the deepest gravitational potentials,
galaxies at significantly larger distances are likely to be
virialized and therefore suitable for mass determinations.  In addition
several of the clusters contained in the Biviano et al.\
study are cool systems, for which this work implies that
unvirialized galaxies will be included within 0.75$h^{-1}$ Mpc.

Escalera et al.\ (1994) calculate the total masses of their clusters by
summing the masses of the individual subclusters, and conclude
that substructure does not greatly affect total cluster masses.  In the
case of irregular systems like A548 and A2151, this may be correct (if
the subclusters have not yet interacted and
remain unperturbed).  In regular galaxy clusters like those with dominant
central galaxies, any detected
substructure, {\it especially} subclusters with a small fraction of the
total mass of the system, are likely to be severely disrupted during their
interactions with the ``host subcluster.''  If these subclusters are
excluded from the estimate of the total cluster mass, the
Escalera et al.\ results are consistent
with those presented here.

No matter how carefully one applies
substructure corrections to datasets and how rigorously
one determines $r_{vir}$,
dynamical mass estimators based on optical data are subject to potentially
large uncertainties.  Merritt (1987) pointed out that the value of a
virial-type mass estimate may vary by factors of hundreds if the shape
of the gravitational potential is unknown, as is the case in most nearby
clusters.  This work is supported by the simulations results of Carlberg \&
Dubinski (1991),
who find that the ``mass-traces-light'' assumption commonly
used to evaluate optical dynamical masses may greatly underestimate the true
mass of the system if the optical velocity dispersions are biased.  X-ray
mass estimation does not suffer from these uncertainties.  It is gratifying
to find that in those cases where high quality X-ray data are available, and
where large optical datasets make objective identification of substructure
straightforward, optical and X-ray mass estimators tend to agree
(Davis et al.\ 1995; Mushotzky 1995).  This result suggests that the
assumption of mass traces light is probably reliable, as is consistent with
preliminary results from gravitational lensing observations (Tyson 1995;
Tyson \& Fischer 1995).

\acknowledgements

It is a pleasure to acknowledge Gus Evrard, Chris Metzler and Keith Ashman
for their advice and suggestions on the physics of mass determinations and
cluster formation.  I am especially grateful to Gus for pointing out the
importance of using a cluster radius defined by the X-ray temperature.
It is a sincere pleasure to thank the many observers responsible for
collecting galaxy redshifts for the last ten years, especially John Huchra,
Margaret Geller,
Ann Zabludoff, Alan Dressler, Steve Shectman, Eliot Malumuth, Bill Oegerle
and John Hill.  Their dedication made this analysis possible.
I'd also like to thank Eliot Malumuth, Alberto Conti
and Sergei Shandarin for their suggestions.  Mike West's prompt
referee'ing was much appreciated.
This work was supported by
NSF EPSCoR grant No.\ OSR-9255223 to the University of Kansas.

{\small
\begin{table}
\caption{Kinematical and Dynamical Quantities for the Limited Cluster Sample}
\begin{tabular}{lccccccccc} \tableline \tableline
&$S_{BI}$&$S_{BI}^{\prime}$&$<r _{\bot}>$&$<r _{\bot}>^{\prime}$
&M$_{PME}$&
M$_{PME}^{\prime}$&r$_{vir}$&M$_p (<r_{vir})$&M$_p ^{\prime}(<r_{vir})$\\
&km s$^{-1}$&km s$^{-1}$&$h^{-1}$kpc&$h^{-1}$kpc&10$^{14}$ M$_{\odot}$
&10$^{14}$ M$_{\odot}$&$h^{-1}$kpc&10$^{14}$ M$_{\odot}$&
10$^{14}$ M$_{\odot}$ \\
\tableline
A85&810$^{+76}_{-80}$&810$^{+76}_{-80}$&773&773
&12.6$^{+3.2}_{-3.3}$&12.6$^{+3.2}_{-3.3}$&1330 &11.8$^{+2.7}_{-3.9}$&
11.8$^{+2.7}_{-3.9}$\\
A119&862$^{+165}_{-140}$&1036$^{+214}_{-221}$&551&271&
 8.7$^{+1.7}_{-2.5}$ &
4.9$^{+0.4}_{-0.4}$&1169&6.8$^{+2.0}_{-1.7}$&4.9$^{+0.5}_{-0.5}$ \\
A193&726$^{+130}_{-108}$&515$^{+176}_{-153}$&375&217&
5.2$^{+2.6}_{-1.6}$&
1.2$^{+0.5}_{-0.5}$&1061 &5.2$^{+1.8}_{-1.5}$&1.2$^{+0.5}_{-0.6}$\\
A194&530$^{+149}_{-107}$&470$^{+98}_{-78}$&536&420&
8.3$^{+1.1}_{-1.1}$&
3.6$^{+2.1}_{-2.0}$&732 &1.9$^{+0.4}_{-0.6}$&1.9$^{+0.4}_{-0.6}$\\
A399&1183$^{+126}_{-108}$&1224$^{+131}_{-116}$&782&677&
23.4$^{+3.1}_{-3.1}$&
21.6$^{+5.0}_{-3.7}$&1268&19.7$^{+4.0}_{-4.1}$&19.9$^{+6.4}_{-4.1}$\\
A401&1141$^{+132}_{-101}$&785$^{+111}_{-81}$&678&732&
19.4$^{+3.3}_{-4.9}$&
12.3$^{+4.2}_{-4.4}$&1518&19.5$^{+3.6}_{-3.5}$&12.2$^{+4.0}_{-3.3}$\\
A426&1262$^{+171}_{-132}$&1262$^{+171}_{-132}$&425&425&
17.2$^{+10.2}_{-5.1}$&
17.2$^{+10.2}_{-5.1}$&1299&13.6$^{+1.8}_{-2.8}$&13.6$^{+1.8}_{-2.8}$\\
A496&741$^{+96}_{-83}$&533$^{+86}_{-76}$&454&376&
6.1$^{+2.1}_{-1.7}$&
2.9$^{+1.4}_{-1.3}$&1035&5.7$^{+2.2}_{-0.9}$&2.3$^{+0.5}_{-0.5}$\\
A754&719$^{+143}_{-110}$&1079$^{+234}_{-243}$&711&823&
10.3$^{+1.5}_{-1.3}$&
15.1$^{+4.8}_{-6.0}$&1527&10.3$^{+1.5}_{-1.3}$&15.1$^{+4.8}_{-6.0}$\\
A1060&630$^{+66}_{-56}$&710$^{+78}_{-78}$&510&316&
3.4$^{+0.4}_{-0.6}$&
2.3$^{+0.2}_{-0.3}$&940&3.1$^{+0.6}_{-0.3}$&2.8$^{+0.4}_{-0.2}$\\
A1644&919$^{+156}_{-114}$&921$^{168}_{-141}$&743&710&
17.1$^{+5.5}_{-7.2}$&
16.9$^{+4.6}_{-4.5}$&1048&14.6$^{+5.2}_{-5.0}$&14.6$^{+5.4}_{-5.1}$\\
A1736&955$^{+107}_{-114}$&528$^{+136}_{-87}$&668&529&
13.3$^{+3.4}_{-3.5}$&
3.1$^{+2.2}_{-1.8}$&1110&9.7$^{+2.4}_{-1.9}$&3.3$^{+2.4}_{-2.1}$\\
A1795&834$^{+142}_{-119}$&912$^{+192}_{-129}$&558&445&
9.5$^{+1.8}_{-1.5}$&
9.4$^{+2.5}_{-1.4}$&1225&9.7$^{+1.0}_{-1.6}$&9.6$^{+2.3}_{-2.0}$\\
A1809&782$^{+148}_{-125}$&851$^{+142}_{-154}$&473&391&
6.5$^{+1.7}_{-1.9}$&
7.2$^{+1.8}_{-1.2}$&--- &---&---\\
A1983&646$^{+184}_{-129}$&532 &650&775&
10.2$^{+5.1}_{-2.6}$ &
3.1$^{+1.4}_{-1.3}$&---&---&---\\
A2052&1404$^{+401}_{-348}$&714$^{+143}_{-148}$&553&270&
50.7$^{+14.3}_{-10.8}$&
3.4$^{+0.6}_{-0.5}$&954&3.7$^{+1.4}_{-1.5}$&2.7$^{+0.6}_{-0.4}$\\
A2063&827$^{+148}_{-119}$&706$^{+117}_{-109}$&459&360&
10.2$^{+5.3}_{-2.8}$&
3.9$^{+0.5}_{-0.4}$&954&5.3$^{+1.1}_{-1.0}$&3.6$^{+0.1}_{-0.3}$\\
A2107&684$^{+126}_{-104}$&577$^{+177}_{-127}$&396&305&
4.4$^{+0.8}_{-0.6}$&
1.8$^{+0.7}_{-0.5}$&1061&4.4$^{+0.7}_{-0.6}$&1.8$^{+0.7}_{-0.5}$\\
A2124&872$^{+151}_{-114}$&906$^{+135}_{-146}$&482&263&
7.2$^{+2.2}_{-2.0}$&
4.1$^{+0.4}_{-0.6}$&---&---&---\\
A2199&829$^{+124}_{-118}$&829$^{+124}_{-118}$&444&444&
6.6$^{+1.4}_{-0.9}$&
6.6$^{+1.4}_{-0.9}$&---&---&---\\
A2634&1077$^{+212}_{-152}$&824$^{+142}_{-133}$&653&506&
32.0$^{+7.5}_{-5.1}$&
10.6$^{+1.1}_{-1.1}$&954&17.0$^{+7.5}_{-5.2}$&7.5$^{+3.9}_{-1.7}$\\
A2670&1037$^{+109}_{-81}$&786$^{+239}_{-203}$&512&505&
9.3$^{+1.1}_{-1.0}$&
7.1$^{+1.8}_{-1.4}$&1022&8.7$^{+1.4}_{-0.9}$&7.0$^{+1.7}_{-2.8}$\\
A3558&923$^{+120}_{-101}$&781$^{+111}_{-98}$&458&377&
9.9$^{+2.5}_{-1.9}$&
5.6$^{+0.9}_{-1.3}$&1009&9.9$^{+2.5}_{-1.8}$&5.6$^{+1.0}_{-1.3}$\\
0107-46&1032$^{+125}_{-108}$&1034$^{+130}_{-115}$&420&348&
12.4$^{+1.0}_{-0.9}$&11.1$^{+2.3}_{-2.2}$&---&---&---\\
1842-63&522$^{+98}_{-82}$&565$^{+138}_{-117}$&695&123&
5.4$^{+3.0}_{-1.4}$&
1.0$^{+0.4}_{-0.3}$&612&2.3$^{+1.0}_{-0.9}$&1.0$^{+0.4}_{-0.3}$\\
\tableline
\end{tabular}
\end{table}
}

\newpage

\begin{center}
{\bf Figure Captions}
\end{center}

Figure 1:  Distributions of velocity dispersion, mean distance of cluster
galaxies from the dynamical centroid, and projected masses for the limited
cluster sample, before and after the correction for substructure.

Figure 2:  The cumulative distribution function of cluster masses.  The
dotted line is the CDF for masses uncorrected for substructure; the
solid line is the CDF for corrected masses.

%

\end{document}